\documentclass[10pt,conference,letterpaper]{IEEEtran}

\usepackage[bookmarks]{hyperref}
\usepackage[dvipsnames]{xcolor}
\hypersetup{
   colorlinks,
   citecolor=Maroon,
   linkcolor=Maroon,
   urlcolor=blue
}

\usepackage{times,amsmath,amssymb,epsfig}
\usepackage{algpseudocode}
\usepackage{algorithm}
\usepackage{balance}
\usepackage{tikz}
\usepackage{comment}
\usepackage{url}
\usepackage{subcaption}

\newcommand{\BMI}{M2}

\title{\BMI: Malleable Metal as a Service}
\author{\IEEEauthorblockN{
Apoorve Mohan\IEEEauthorrefmark{1},
Ata Turk\IEEEauthorrefmark{2}, 
Ravi S. Gudimetla\IEEEauthorrefmark{3},
Sahil Tikale\IEEEauthorrefmark{2},
Jason Hennesey\IEEEauthorrefmark{2},\\
Ugur Kaynar\IEEEauthorrefmark{2},
Gene Cooperman\IEEEauthorrefmark{1},
Peter Desnoyers\IEEEauthorrefmark{1},
and
Orran Krieger\IEEEauthorrefmark{2}
}
\vspace{1.6mm}\\
\IEEEauthorblockA{
\IEEEauthorrefmark{1}Northeastern University,
\IEEEauthorrefmark{2}Boston University,
\IEEEauthorrefmark{3}Red Hat Inc.,
}}

\begin{document}
\maketitle
\begin{abstract} 
  Existing bare-metal cloud services that provide users with 
  physical nodes have a number of serious disadvantage over
  their virtual alternatives, including slow provisioning times,
  difficulty for users to release nodes and then reuse them to handle
  changes in demand, and poor tolerance to failures.  We introduce
  \BMI, a bare-metal cloud service that uses network-mounted boot
  drives to overcome these disadvantages. We describe the architecture
  and implementation of \BMI\ and compare its agility, scalability and
  performance to existing systems.  We show that \BMI\ can reduce
  provisioning time by over 50\% while offering richer functionality,
  and comparable run time performance with respect to tools
  that provision images into local disks.  \BMI\ is open source and
  available at \url{https://github.com/CCI-MOC/ims}.
\end{abstract}

\section{Introduction}
\label{sec:intro}

Although virtualized cloud services can satisfy the requirements of many applications, some applications still require physical (i.e., bare-metal) nodes. Examples include performance or security sensitive applications that cannot tolerate the overhead, unpredictability, and large trusted computing base of complex virtualized   cloud services~\cite{Facebook:2011, Iosup:2011, Ristenpart:2009}, or applications that need direct and exclusive access to hardware components that are difficult to virtualize (e.g., InfiniBand~\cite{softlayer_infiniband}, RAID~\cite{Barker:2010}, FPGAs~\cite{FPGA:2014}, GPUs~\cite{Maurice2014}, etc.). 

Cloud vendors have developed application-specific solutions dedicated to some of these use cases (e.g., Amazon HPC cloud~\cite{aws_HPC}, Amazon GPU nodes~\cite{aws_GPU}, Cirrascale deep learning cloud~\cite{Cirra_GPU}, etc.).  However, these compartmentalized solutions lead to cloud silos, reducing the flexibility of the cloud to move resources between different users as demand warrants. Moreover, it is impossible to cover all bare-metal use cases with dedicated solutions; consider, for example, researchers that want to develop their own bare-metal operating system~\cite{schatzberg2016ebbrt}, or cloud developers that need to test software on environments identical to the eventual production environments. 

The demand for bare-metal clouds has resulted in an increasing number of offerings such as IBM~\cite{Softlayer:2015}, Rackspace~\cite{RackSpace:2015}, and Internap~\cite{Internap:2015}. These bare-metal cloud solutions install the tenant's operating system and application into the server's local disks.  This installation process incurs long startup delays (tens of minutes to hours) and high networking costs to copy large disk images.  Moreover, because user state is local to the server, these solutions lack rich functionality of virtual solutions including checkpointing/cloning of images, releasing and re-aquiring nodes to match demand, and fast recovery from node failures.

A number of industry and research projects have attacked the performance and functionality challenges of provisioning bare-metal nodes~\cite{Ironic:2015,MAAS:2015,anderson_automatic_2006, OpenCrowbar:2015, Razor:2015, Omote:2015}; automating the bare-metal provisioning process, reducing the management overhead of the cloud provider, and improving the performance of copying the image to the server's disk.  For example, Omote \hbox{et al.}~\cite{Omote:2015} proposed a lazy copy approach that copies the OS image in the background after the operating system is booted using a remote disk.  While sophisticated techniques like this can reduce some of the user visible provisioning time, all these approaches end up eventually transferring the boot image to the local disk, and hence still incur overhead to copy the image and have the functionality problems discussed above. 

We present {\em \BMI}, a provisioning tool for bare-metal clouds that addresses the challenges described above. Similar to virtualized cloud services, \BMI\ serves user images that contain the operating system (OS) and applications from remote-mounted boot drives. 
\BMI\ relies on a fast and reliable distributed storage system (CEPH~\cite{CEPH2006, CEPH2017} in our implementation) for hosting images of provisioned bare-metal instances and a network isolation service (HIL~\cite{hil} in our implementation) for isolating tenants in the cloud. 

Key contributions of this work include:
\begin{enumerate}
\item The definition of \BMI\, a general purpose architecture of a bare-metal cloud provisioning system that exploits remote storage\footnote{Previous provisioning systems exploited remote storage in special purpose environments, like HPC clusters, where all nodes boot the same kernel.} and allows users to: 
\begin{itemize}
     \item rapidly release and then acquire nodes to handle fluctuation in demand, 
     \item rapidly recover from failed nodes by booting another node with the disk,
     \item snapshot and clone disk images,
     \end{itemize}
\item An implementation and analysis that demonstrates that: 
	\begin{itemize}
        \item it is possible to provision and deploy bare-metal systems with overheads similar to deploying virtual machines, 
        \item performance of the \BMI-provisioned servers is similar to those provisioned to local disks.\footnote{As the focus of this work is to improve the provisioning time \BMI\ only network-mounts boot drives that hosts the OS and applications. Data drives are still hosted on the local disks.} 
    \end{itemize}
\end{enumerate}

The remainder of this paper is organized as follows.
We provide related work in Section~\ref{sec:bg}.
The design and architecture of \BMI\ are presented in Section~\ref{sec:design} and Section~\ref{sec:arch}, respectively.
We evaluate performance, scalability, and
usability of \BMI\ in Section~\ref{sec:eval}. 
We discuss future directions for \BMI\ in Section~\ref{sec:fw} and conclude in Section~\ref{sec:con}.

\section{Related Work}
\label{sec:bg}

In this section we review existing bare-metal provisioning approaches considering their fitness to support on-demand bare-metal IaaS offerings.  
We can broadly classify provisioning approaches into two as {\em diskful}
and {\em diskless} provisioning systems, based on where the image is
hosted once a bare-metal instance is provisioned.

{\em Diskful Provisioning Systems:}
These systems persist the provisioning image to the local disks of the
bare-metal systems. The standard provisioning tools used in many bare-metal deployments are diskful. A rich set of open source and commercial provisioning products such as Emulab~\cite{anderson_automatic_2006}, OpenStack Ironic~\cite{Ironic:2015}, Crowbar~\cite{OpenCrowbar:2015}, Cannonical Metal-as-a-Service (MaaS)~\cite{MAAS:2015}, Razor~\cite{Razor:2015}, and Cobbler~\cite{Cobbler:2015} are available for automated diskful provisioning of bare-metal systems. 
Chandrasekar and Gibson~\cite{Chandrasekar2014} provide a comparative analysis of commonly used diskful provisioning systems.

Diskful provisioning systems can be further divided into two types.
The first type of solutions automate the manual installation process of the OS and desired applications to the
local disks (e.g., Foreman~\cite{foreman}).
As they follow a step by step installation process,
these solutions generally take the longest to provision.
The second type of solutions copy a
pre-installed image, containing the operating system and applications,
onto the local disk over the network (e.g., OpenStack  Ironic~\cite{openstack-ironic}).
The size of such pre-installed images can be tens of GBs.
Transferring them can overwhelm the network and persisting them to local disks still requires hundreds of seconds assuming standard HDDs are used.
Both solutions have a lower bound on the time before a node is ready to use due to the need to reboot twice\footnote{Rebooting modern
datacenter servers can take as long as 5 minutes~\cite{slowboot}}, once via PXE to enter the installer, and another to boot into the freshly installed system.

When using diskful systems, repurposing a bare metal system requires
formating the local disks and then installing/copying the new system. If
saving the existing disk state is desirable, the contents of the disk
has to be copied away, which again requires hundreds of seconds and
further increases the re-provisioning cost.

In general, diskfull systems and the automation tools that employ these provisioning systems (e.g., Ironic, MaaS, Foreman) are designed for setting up long running bare-metal systems. They consider the high startup delays tolerable assuming that the servers they provision will have long operational lifetimes. 
To support fast provisioning and reducing boot time of diskfull systems Omote et al.~\cite{Omote:2015} propose BMCast, an OS deployment system with a special purpose de-virtualizable Virtual Machine Manager that supports OS-transparent quick startup of bare-metal instances.

Beyond the cost of transfering images to the local disk, a fundamental problem with all diskful provisioing systems is that any modifications to the image are stored on the disks attached to the physical note allocated to a user.  This means, for example, that a user cannot easily release and re-acquire nodes to match their needs, since any state on the local disk is lost when the user releases the nodes.  Some of the rich functionality users take for granted in virtualized clouds is also not available in these environments.  For example, a user cannot snapshot the disk of a physical node and then clone it to boot additional nodes.  Perhaps most importantly, if a physical node fails, the user cannot easily start up another node from the same disk image or use the disk to diagnose the failure; any state local to that disks node is inaccessible as long as the node is down.  Moroever, 
in diskfull systems the local disk hosted boot drives become a single point of failure. If the disk containing the boot drives fails, the bare-metal node becomes unusable until the disk is fixed or replaced and data recovery can be daunting in such cases. 

{\em Diskless Provisioning Systems:} 
These systems keep the provisioning image resident on a network
accessible remote logical disk that appears as a local disk to bare-metal
systems. This method of provisioning historically has been used with
diskless workstations~\cite{klimenko1999technique, sposato2002method, haun2004providing} and HPC systems~\cite{salah2011towards, guler2002advantages, daly2007base} to boot multiple nodes from a single image. Furthermore, diskless provisioning is heavily used in virtualized systems~\cite{lewis2005virtual, cully2008remus, nelson2005fast}.
Interestingly, diskless provisioning is not being used in cloud deployments for bare metal provisioning and there are no tools or studies that combine diskless provisioning with image management capabilities to support bare-metal provisioning and servicing of multiple images owned by multiple users. To our knowledge, \BMI\ is the first effort in this direction.

\section{\BMI\ Design}
\label{sec:design}

When designing \BMI\ we first listed the set of features we wanted to have in order to support an on-demand bare-metal cloud service. These features are:

\begin{itemize}

\item{{\em Rapid provisioning:}} A bare-metal cloud service has to offer
on-demand bare-metal servers with minimum startup overhead so that even
short lived deployments with life-span of a few hours can use bare-metal
servers efficiently. If servers take tens of minutes to deploy, the
``effective'' utilization of the cloud will decrease.

\item{{\em Rapid snapshotting, re-purposing, reprovisioning:}} Abilities
for quickly snapshotting the OS and applications, releasing a server when
unused, and being able to quickly provision a server using a previous
snapshot are critical for time-multiplexing bare-metal servers across
many users. These features also enable the service to offer ``elasticity''
to the applications it hosts.

\item{{\em Rapid cloning:}} The ability to rapidly stand up a large
number of servers concurrently using the same saved image is a common
request in infrastructure as a service clouds. This feature enables easy
deployment of parallel/distributed applications and scalability.

\item{{\em Support for multi-tenancy:}} Existing provisioning tools
assume that they are available to just the administrator of the hardware
and all of the hardware available in the system is managed by the same
entity. However, in a bare-metal cloud service, the provisioning system
has to ensure performance isolation and security across its users even
during provisioning.
\end{itemize}

Given the above list of desirable features, we made a set of design decisions for \BMI. In order to offer rapid provisioning, we opted to use diskless provisioning mechanisms. 
Using these mechanisms \BMI\ does not need to copy the entire image to the bare-metal server and it can save the overhead to install images and applications to local disks once a cloud image is prepared. \BMI\ can rapidly start running applications by only fetching the necessary OS and application libraries before start-up and further required packages will be fetched on-demand as they are used. Note that the standardization of technologies such as iSCSI~\cite{rfc7143} has allowed diskless provisioning to be used with commodity servers and clients over any layer-3 network.

We also decided to network boot the nodes from images residing on a distributed storage. \BMI\ can service the images from centralized high performance storage systems using multiple disks to improve boot time. Furthermore, many modern distributed storage systems support capabilities such as {\em copy-on-write (COW)}, {\em de-duplication} and {\em linked-cloning}~\cite{LinkedClone:2017}, which are beneficial for capabilities such as rapid cloning and snapshotting.

We note that in diskless provisioning clients are always
dependent on uninterrupted access to the centralized storage, and as
the number of clients increase, the storage infrastructure has to be
adequately scaled to support the increasing load. Network connectivity and
availability also plays a critical role in the performance of diskless
provisioning systems.
However, with the advancements in faster, cheaper and redundant networking (e.g., Clos
networks~\cite{clos}) and storage solutions (e.g., Solid State Drives), datacenter applications increasingly lean towards disaggragating
storage services to make full use of the capacity of their
infrastructures~\cite{klimovic:2016}. Diskless provisioning approaches are
very much aligned with this trend. 

\begin{figure}[t]
\vspace{-1ex}
  \centering
   \includegraphics[width=0.99\columnwidth]{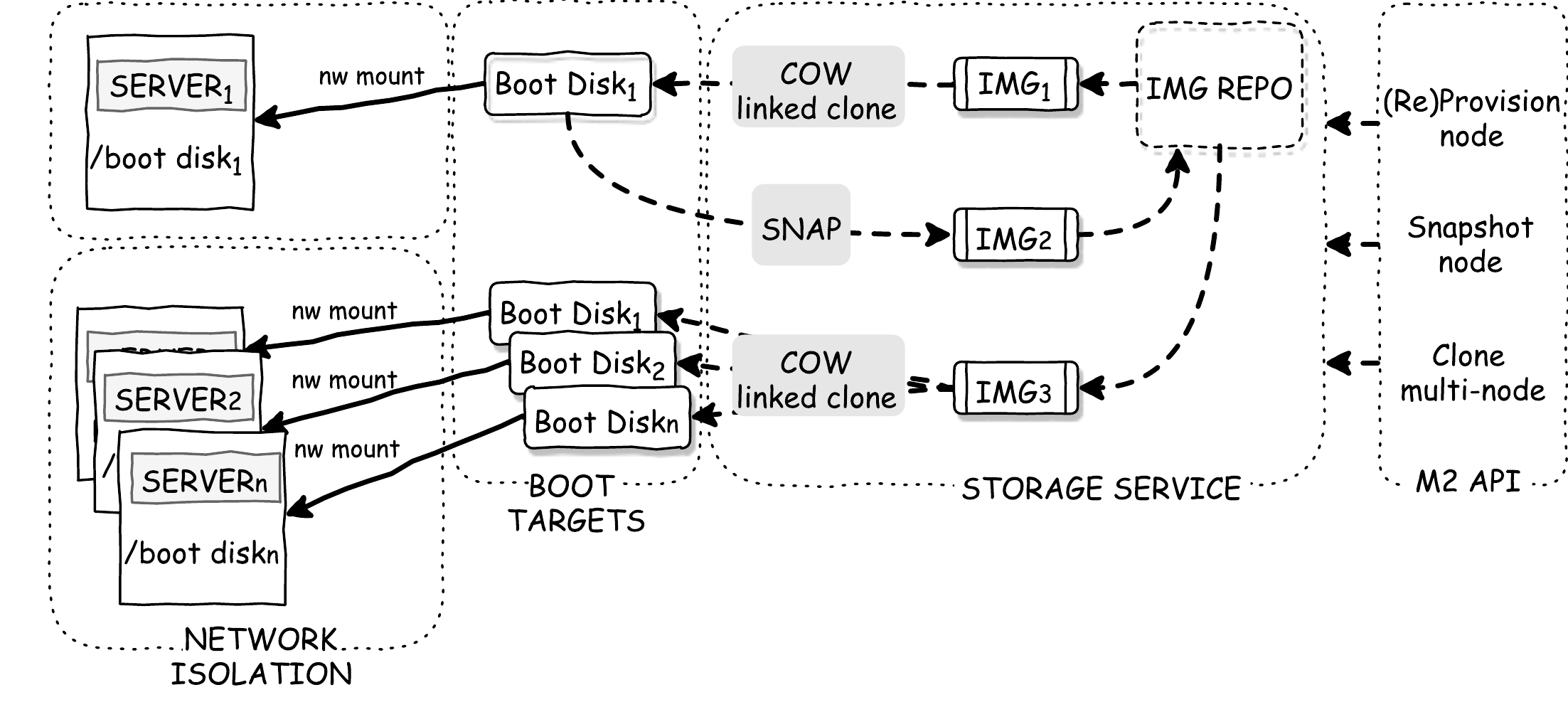}
  \vspace{-1ex}
  \caption{\BMI\ mockup design}
  \label{fig:bmi-concept}
  \vspace{-1ex}
\end{figure}

Figure~\ref{fig:bmi-concept} presents the conceptual design for \BMI\ and some of the functionalities it offers. As seen in the figure, \BMI\ stores images (user or \BMI\ provided) in an image repository. When a provisioning API call is made for a bare-metal node with a given image, a linked-clone of that image is created, followed by network isolation of the requested bare-metal node and mounting of the clone on the bare-metal node. When the node network-boots, it only fetches the parts of the image it uses, which significantly reduces the provisioning time.
\BMI\ also supports provisioning multiple nodes in parallel from a single image by simply performing parallel provisioning calls.   

As seen in Figure~\ref{fig:bmi-concept}, disk snapshotting API call of \BMI\ enables users to create checkpoints/restore-points by saving the current state of the image to the repository and tagging the saved image with a unique identifier. Using linked-cloning and COW, \BMI\ can offer rapid snapshotting. 

\begin{figure}[t]
\vspace{-1ex}
  \centering
   \includegraphics[width=0.99\columnwidth]{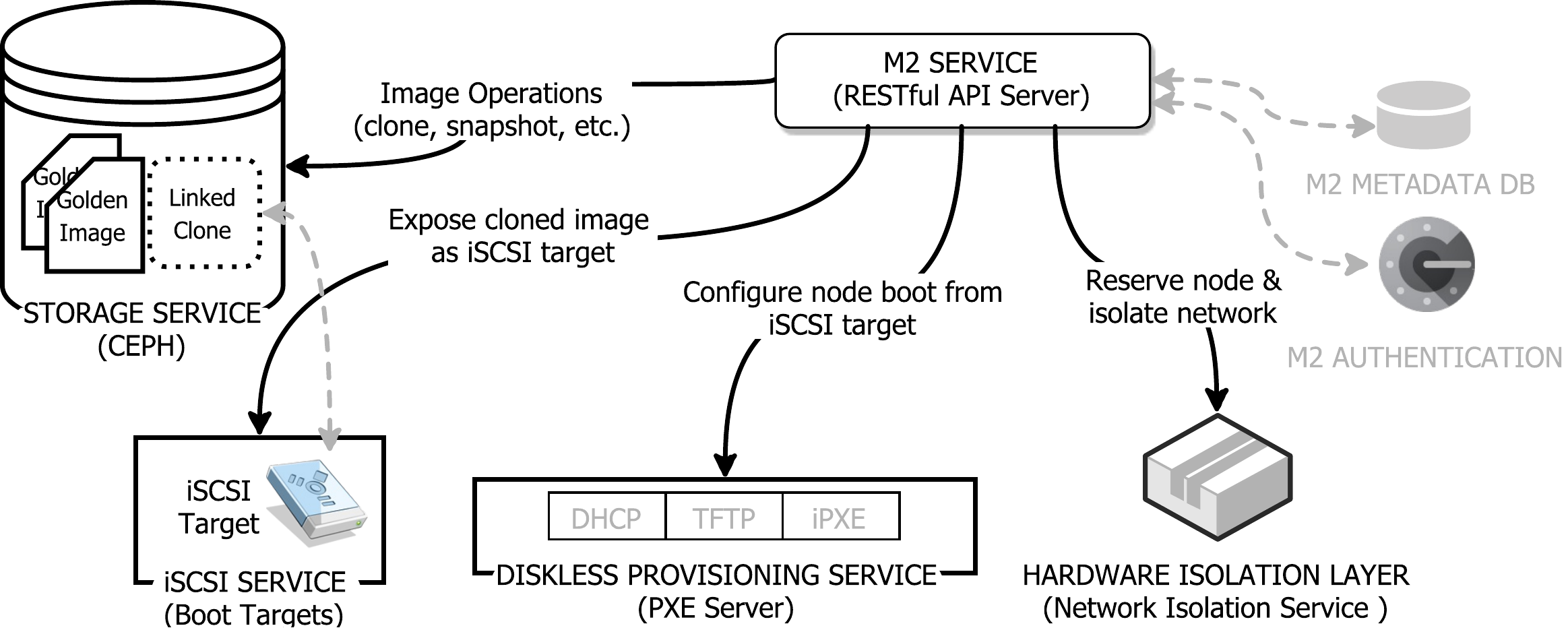}
  \vspace{-1ex}
  \caption{\BMI\ components and architecture}
  \label{fig:bmi-impl}
  \vspace{-1ex}
\end{figure}

\section{\BMI\ Architecture}
\label{sec:arch}

In this section we discuss our implementation for \BMI. There are five major components in \BMI: (i) API Server, (ii) Storage Service (Ceph), (iii) iSCSI Service (TGT Server), (iv) Diskless Provisioning Service (PXE Server), and (v) Network Isolation Service. Figure~\ref{fig:bmi-impl} displays these five components. \BMI\ follows a driver based approach and provides an abstraction for each component.
This allows system administrators to replace the solution used for any of these components.
For example, in the current implementation Ceph~\cite{weil2007ceph} is used as the storage service, but it is possible to replace it with any other storage service that supports COW like the network-based Lustre~\cite{lustre} or even local systems like ZFS~\cite{zfs} or Linux's LVM~\cite{lvm}.

\textbf{Storage Service:} Storage Service provides a data store for the cloud images and exposes API's to rapidly clone and snapshot existing images. In our implementation we use Ceph as our storage solution. Ceph is an open source storage platform that implements a highly reliable and scalable object storage on a distributed cluster~\cite{weil2007ceph}. It exposes various interfaces for object, block and file level storage~\cite{CEPH2017}. We used the block storage interface provided by Ceph aka the Reliable Autonomic Distributed Object Store Block Device (RADOS Block Device or RBD) to store the cloud images using the {\em librados} API. {\em librados} also exposes functionalities such as cloning and snapshotting to manage the RBD based cloud images.
Ceph provides data store and image management capabilities like snapshotting and cloning and offers good read performance~\cite{cephperf}, which helps in achieving lower latency when \BMI\ tries to fetch the disk blocks on-demand~\cite{iscsistorageinterconnect}.

\textbf{iSCSI and Diskless Provisioning Services:} Our implementation of diskless provisioning is based on network booting bare-metal nodes from RBD based cloud images (stored in CEPH) that are exposed as iSCSI targets. We used the Linux SCSI Target Framework (TGT)~\cite{tomonori2006tgt} to expose the RBD based cloud images as iSCSI targets.
As TGT is a user-space implementation, no extra kernel code is required which improves its compatibility with modified Linux kernels. Being a user-space server also supports multi-tenancy within \BMI\ and defense in depth by enabling the iSCSI server to be run in a Linux container~\cite{docker}, which can permit a single physical node to serve different tenants on different networks without exposing all the iSCSI endpoints to all tenants.
TGT also provides native support for RBD based images, freeing \BMI\ from managing additional mapping state.
The current implementation of \BMI\ does not have support for iSCSI multipathing or special iSCSI hardware, however, in the upcoming release of \BMI\ we are working to provide support for both scenarios (see \autoref{sec:future-scaling}).

Preboot eXecution Environment (PXE) specification provides a standardization for a client-server model for booting nodes over the network using Dynamic Host Configuration Protocol (DHCP) and Trivial File Transfer Protocol (TFTP). Although, PXE provides specifications to network boot a node from various targets (HTTP, iSCSI, AOE etc.), it is up to the NIC manufacturer to implement the support to network boot from a particular target into the NIC firmware. As mentioned earlier, \BMI\ uses an iSCSI based approach to network boot the bare-metal nodes. To ensure that \BMI\ can provision any bare-metal node irrespective of the NIC firmware capabilities of that node, \BMI\ first chainloads~\cite{chain} into iPXE, which eventually network boots the bare-metal nodes from the exposed iSCSI target. iPXE is an open-source implementation of network booting firmware that provides all the features mentioned in the PXE specifications~\cite{chain}.

The iSCSI Boot Firmware Table (iBFT)~\cite{ibft} gives PXE servers the ability to specify an iSCSI target to which the tenant OS should connect. iBFT makes \BMI\ OS-agnostic, since it eliminates the need for OS-specific parsing and modification of images to configure the identity of the iSCSI server for a given node.

\textbf{\BMI\ API Server:} API Server is a python based RESTful web service that controls the flow between different components of \BMI. Exposed APIs enable users to (de)provision nodes, clone provisioned nodes, create snapshots of provisioned nodes, (de)register users, perform various operations pertaining to images (upload, download, rename, share, list, etc.), list various resources, etc. The API server also maintains a database for various book keeping purposes such as maintaining the mapping between bare-metal nodes and cloud images, user-cloud image mappings, etc. 

While most of the API calls are trivial and trigger various \BMI\ database operations, some of them accounts to interacting with different \BMI\ components --- in particular the APIs pertaining to (re/de)-provisioning, snapshoting  and cloning nodes, and image manipulation. APIs that require interacting with the storage service rely heavily on the performance of the exposed block storage management capabilities. 

The provision API enables users to spawn bare-metal instances from existing cloud images hosted in the storage service\footnote{Cloud images need to be registered and uploaded to \BMI\ before the provision API is invoked}. It accepts the ID defining the node to be provisioned (e.g. MAC address, NIC Number) and the ID of the cloud image to be used for provisioning as arguments. Upon receiving a provision request, the API server interacts with the Storage Service and creates a linked-clone of the cloud image (passed as the argument to the provision call) and exposes it as an iSCSI target. This is followed by the preparation of the PXE and iPXE configuration files by the Diskless Provisioning Service that will be served to the bare-metal node upon its network boot request. This is similar to how virtual machines are provisioned in different IaaS cloud offerings.

\BMI\ exposes a snapshot API that allows users to create checkpoints/tags by saving a deep copy of the existing node state. Users can use these snapshots to revert back to any previous state in case of a failure\footnote{The current implementation of \BMI\ is limited to disk snapshots}. This feature also enables users to manage different configurations of their nodes/clusters. \BMI\ does not expose an explicit clone API. But users can clone an existing node by creating a snapshot of the current node state and provisioning a/multiple new node(s) from that snapshot.

\textbf{Multi-tenancy and Allocations:} For multi-tenancy it is important to segregate each \BMI\ node based on ownership and physically using some network isolation mechanism.
\BMI\ uses Hardware Isolation Layer (HIL)~\cite{hil} for allocations and to achieve multi-tenancy. HIL is a lightweight python-based layer-2 bare-metal isolation framework that orchestrate data center compute resources by controlling the networking infrastructure. It exposes an API that enables users to create isolated groups of compute resources from a hardware resource-pool. HIL is a network-switch agnostic framework that follows a driver-based model\footnote{Currently HIL can manage network isolation for Cisco, Juniper, Dell and Brocade switches.}. HIL is agnostic to the provisioning system running on top of it and is thus our choice for achieving network isolation (multi-tenancy) and allocations.

\section{Experimental Evaluation}
\label{sec:eval}

In this section we evaluate \BMI's speed, scalability and performance. We start by presenting our experimental setup. Then we compare the provisioning time of \BMI\ with that of existing provisioning solutions, present the time taken by different \BMI\ API calls, and analyze the scalability of \BMI. Network overheads associated with using a diskless solutions are also provided and \BMI's impact on the performance of frameworks and applications is analyzed.
We note that in the following experiments, only the boot drives are mounted remotely by \BMI\ as currently \BMI\ focuses on improving boot performance. Whenever data drives are used by applications, those drives are hosted on local disks.

\subsection{Experimental Setup}
\label{sec:exp-setup}
In our experiments we used two different environments. In the first environment, each bare-metal server has two 6-core Intel Xeon E5-2630L CPUs (24~cores with hyperthreading enabled), 300 GB 10K SAS HDDs (two nodes had 1 TB 7.2K SATA HDDs), 128GB RAM and two Intel 82599ES 10 Gbit NICs. A four-node Fujitsu CD10000 Ceph storage cluster with four 10 Gbit external NICs and internal 40 Gbit InfiniBand interconnect is used as the storage server of \BMI\ in this environment. 

In the second environment each bare-metal server has a single Intel 8-core Xeon E5-2650 CPU (2.30GHz, 16 cores with hyperthreading enabled), 64 GB RAM, two 1.8 TB HDDs, and one 10Gbit Ethernet adapter. A ten-node Ceph cluster with a total of 90 spindles and 10GbE internal 40GbE external NICs are used as the storage server of \BMI\ in this second environment. 

For both environments, \BMI's iSCSI and API servers were deployed on a virtual machine with 4 VCPUs and 4 GB RAM. The RHEL 7.1 (i.e., Centos 6.7) operating systems (OS) is installed in cases where an OS installation is performed.

\begin{figure}[t]
\vspace{-1ex}
  \centering
   \includegraphics[width=0.75\columnwidth]{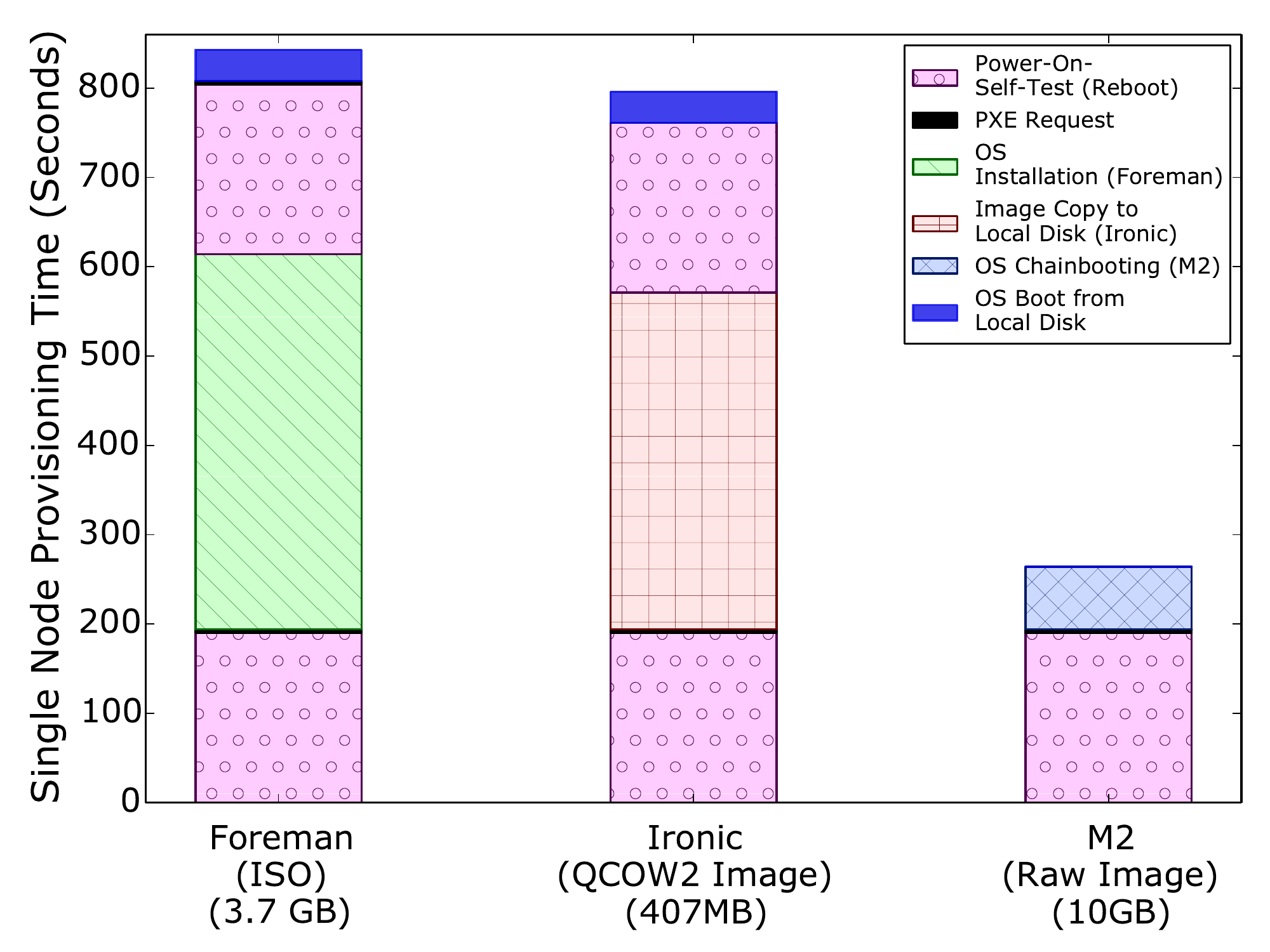}
  \vspace{-1ex}
  \caption[Single node provisioning time comparison between \BMI, Foreman, and OpenStack/Ironic.]{Single node provisioning time comparison between \BMI, Foreman, and OpenStack/Ironic.\footnotemark}
  \label{fig:provision}
  \vspace{-1ex}
\end{figure}

\footnotetext{Bare-metal nodes were provisioned with RHEL 7.1 for all the provisioning systems. The virtual disk size of the image used for both Ironic and \BMI\ was 10 GB. The actual disk size in the case of Ironic was 407 MB whereas for \BMI\ it was 10 GB.}

\subsection{Provisioning Time Comparison} 
\label{sec:provision}

Figure~\ref{fig:provision} presents the time comparison of \BMI\ with Foreman, and OpenStack/Ironic, two widely used provisioning systems, when we provision a single bare-metal server in our first environment with a bare RHEL 7.1 operating system. As seen in the figure, the \BMI\ provisioning time is around five minutes. Note that firmware initialization of these bare-metal servers requires more than three minutes; hence half of the \BMI\ provisioning time is spent in firmware initialization. Both Foreman, and OpenStack/Ironic have to go through firmware initialization phase twice. Furthermore, they have to install or network-transfer the OS to local disk, whereas \BMI\ simply provisions the node out of a remote disk containing the operating system. Due to these advantages, \BMI\ provisions nodes around three times faster than both Foreman and OpenStack/Ironic\footnote{In Figure~\ref{fig:provision}, we do not include the time taken to prepare the provisioning target for the provisioning systems since this is a manual process for Foreman.}.

\begin{figure}[t]
\vspace{-1ex}
  \centering
   \includegraphics[width=0.75\columnwidth]{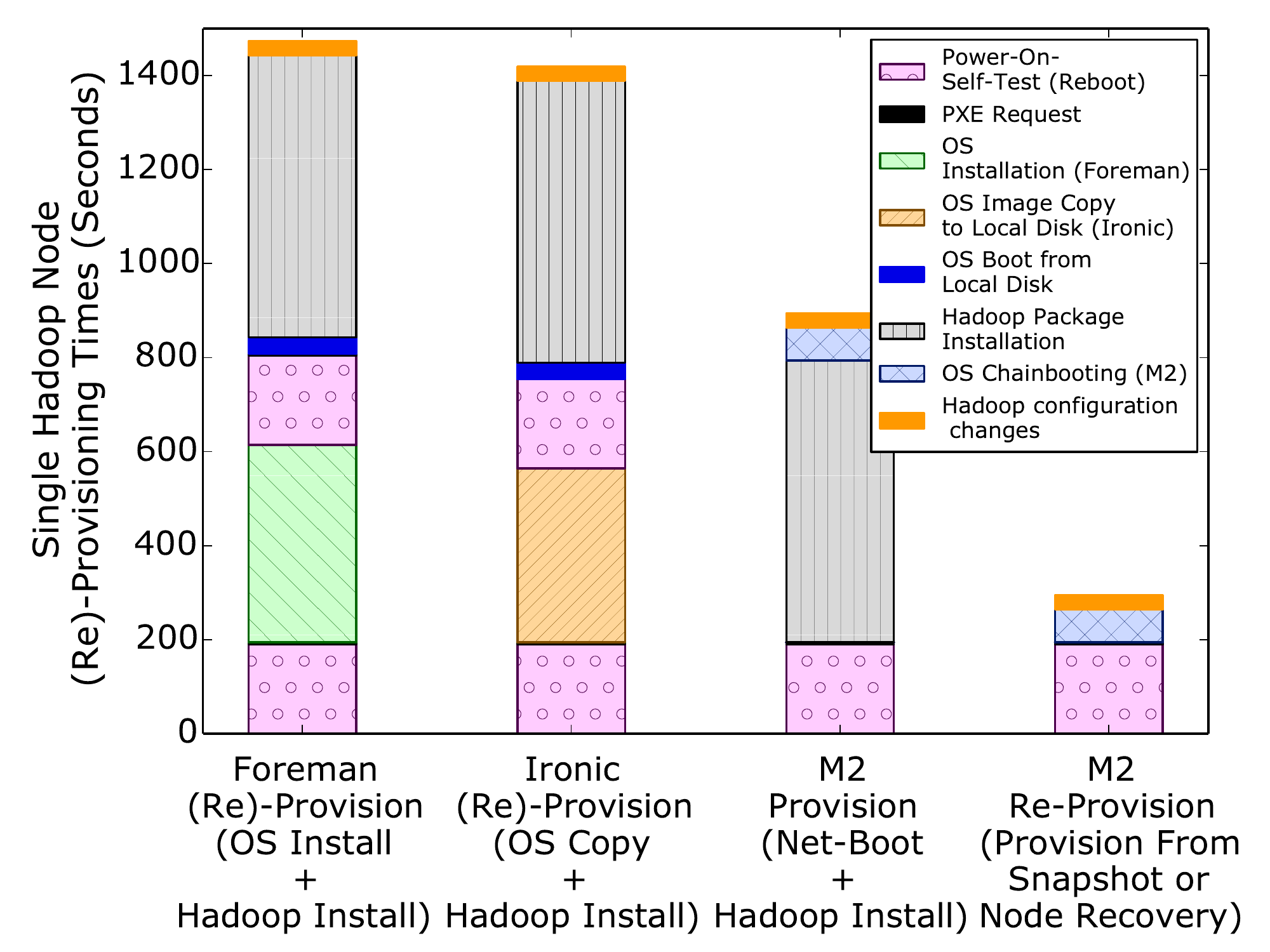}
  \vspace{-1ex}
  \caption{Single Hadoop compute node (re)-provisioning time comparison between \BMI\, Ironic and Foreman.}
  \label{fig:reprovision}
  \vspace{-1ex}
\end{figure}

\subsection{Provisioning Complex Frameworks} 
\label{sec:reprovision}

Provisioning a node for any framework such as Hadoop or SLURM is a complex and time consuming process handled in many-steps. First, the operating system is installed, and then the relevant packages for the framework is installed, which is followed by making configuration changes to the node. The first three bars of Figure~\ref{fig:reprovision} show the total provisioning time for a single Hadoop compute node when using Foreman, Ironic and \BMI. As seen in the figure, \BMI\ can only offer $\sim$40\% improvement during this process as it is dominated by the application installation and configuration. 

Even though installing and configuring frameworks such as Hadoop is a time consuming process, once a single example setup is made, \BMI\ can leverage its snapshotting and cloning mechanism to safe-keep that example and use it for provisioning other framework nodes. As shown in the fourth bar in Figure~\ref{fig:reprovision}, with \BMI, provisioning cloned images that contain desired applications and then doing a final reconfiguration is significantly faster than provisioning nodes from scratch.      

\subsection{Using \BMI\ for Failure Recovery} 
\label{sec:reprovision}

In large data center deployments, node failure is a common phenomenon~\cite{meza2015large, meza2015revisiting, barroso2013datacenter, schroeder2010large}. Recovering from a node failure involves tedious manual operations. If the node is provisioned from the local disk (using Foreman or Ironic), the node becomes unavailable until it is fixed. In addition, if the cause of node failure was disk failure, there is a good chance that all of the user data is lost. In order to re-provision another bare-metal server using Foreman or Ironic as the same Hadoop node, it is required to re-install (or re-copy in the case of Ironic) the operating system and Hadoop packages on the server --- leading to a re-provisioning time similar to the provisioning time. As shown in Figure~\ref{fig:reprovision}, the total time to re-provision a single Hadoop compute node is the same as its provisioning time for Foreman and Ironic.

On the other hand, if this node was provisioned using \BMI, upon failure a new node can be re-provisioned (rebooted) using the image of the failed node that resides in Ceph. The time to re-provision the new node is significantly reduced as there is no requirement to re-install the operating system or any Hadoop packages. As shown in the last bar of Figure~\ref{fig:reprovision}, \BMI\ reduces the re-provisioning time of the nodes by up to 5 times as compared to Foreman or Ironic.

\subsection{Operation Times of Other \BMI\ Calls} 
\label{sec:bmi_overhead}

Table~\ref{tab:bmi-api-overhead} presents the time it takes to perform some of the other \BMI\ API calls that require interaction with the storage service. The time taken by the De-Provision operation constitutes the time for those operations pertaining to HIL (detaching the provisioning network from the node), the iSCSI service (disabling the iSCSI target), the storage service (deleting the image associated with the node to be de-provisioned) and the API server orchestration. The time taken by the Snapshot and Clone Image operations are dominated by the storage operations, which include the time taken to flatten a linked clone in the case of the Snapshot operation and the time taken to create a deep copy of a cloud image in the case of the Clone Image operation.

\begin{table}[t]
\caption{Time required by other \BMI\ operations.} \label{tab:bmi-api-overhead} 
\centering
\begin{tabular}{ l r } 
\BMI\ API Call &  Time (secs) \\ 
\hline
De-Provision & 32.00 \\ 
Snapshot     & 11.65 \\ 
Clone Image  & 7.10  \\
\hline
\end{tabular}
\end{table}

\subsection{Scalability} 
\label{sec:scaling}

\begin{figure}[t]
\vspace{-1ex}
  \centering
   \includegraphics[width=0.75\columnwidth]{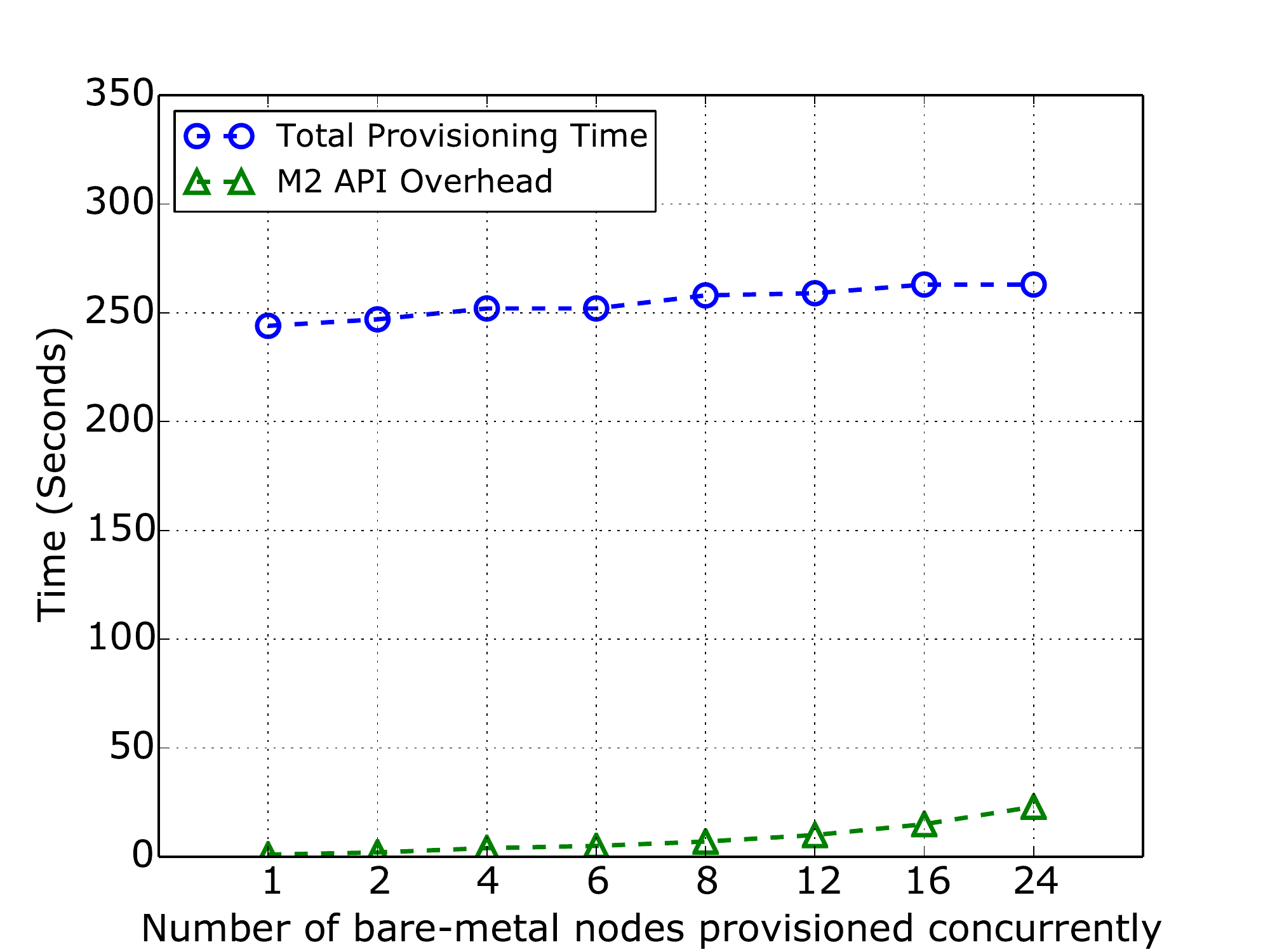}
  \vspace{-1ex}
  \caption{\BMI\ scalability analysis.}
  \label{fig:scaling}
  \vspace{-1ex}
\end{figure}

In Figure~\ref{fig:scaling}, we show the time \BMI\ requires for provisioning multiple nodes in parallel from our second environment. We increase the number of concurrently provisioned nodes from one to~24 and report the time it takes to provision that many nodes (upper blue circle line). As seen in the figure, provisioning 24 nodes takes only around 20 seconds longer than provisioning a single node, indicating that even with modest resource usage \BMI\ is scalable. We observe a slight increase in the \BMI\ overhead (lower green triangle line) since the iSCSI server VM, which has four vCPUs, has to context switch when the number of nodes increase above four.

We note here that the current \BMI\ implementation is totally unoptimized and runs multiple \BMI\ services on a single wimpy VM. We expected to see a significant performance degradation in our scalability analysis as requests on \BMI\ increased but observed that not to be the case. As will be shown in the coming sections, this is due to the fact that only a tiny fraction of the provisioning image is accessed during booting and application runs and the load on the \BMI\ services is comparatively low.   

\subsection{\BMI\ Network Traffic Analysis}
\label{sec:traffic-analysis}

\begin{figure}[t]
\vspace{-1ex}
  \centering
	\begin{subfigure}[t]{0.45\textwidth}
		\centering
		\includegraphics[width=0.75\textwidth]{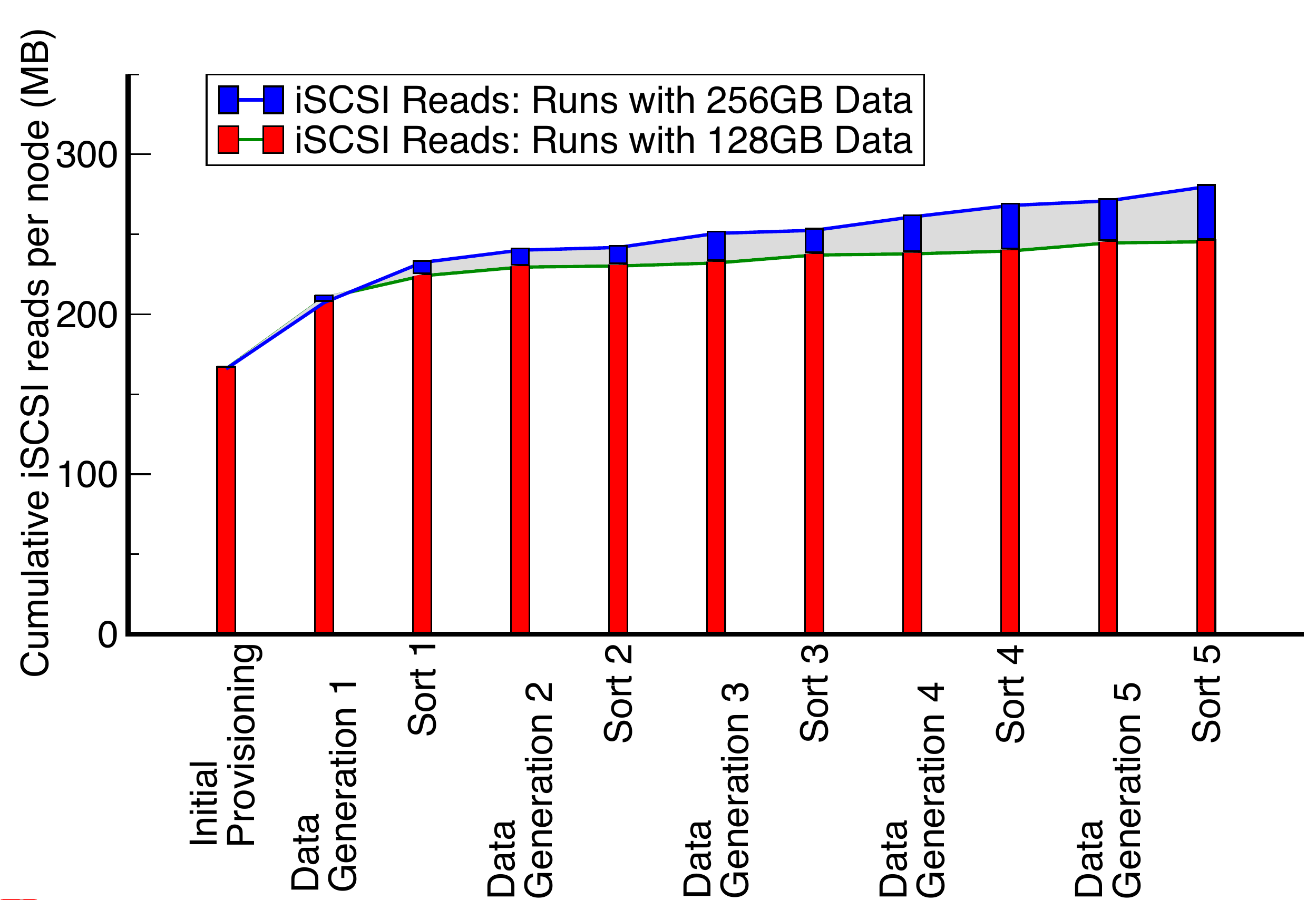}
		\caption{Read Traffic} \label{fig:read-traffic}
	\end{subfigure}
	\begin{subfigure}[t]{0.45\textwidth}
		\centering
		\includegraphics[width=0.75\textwidth]{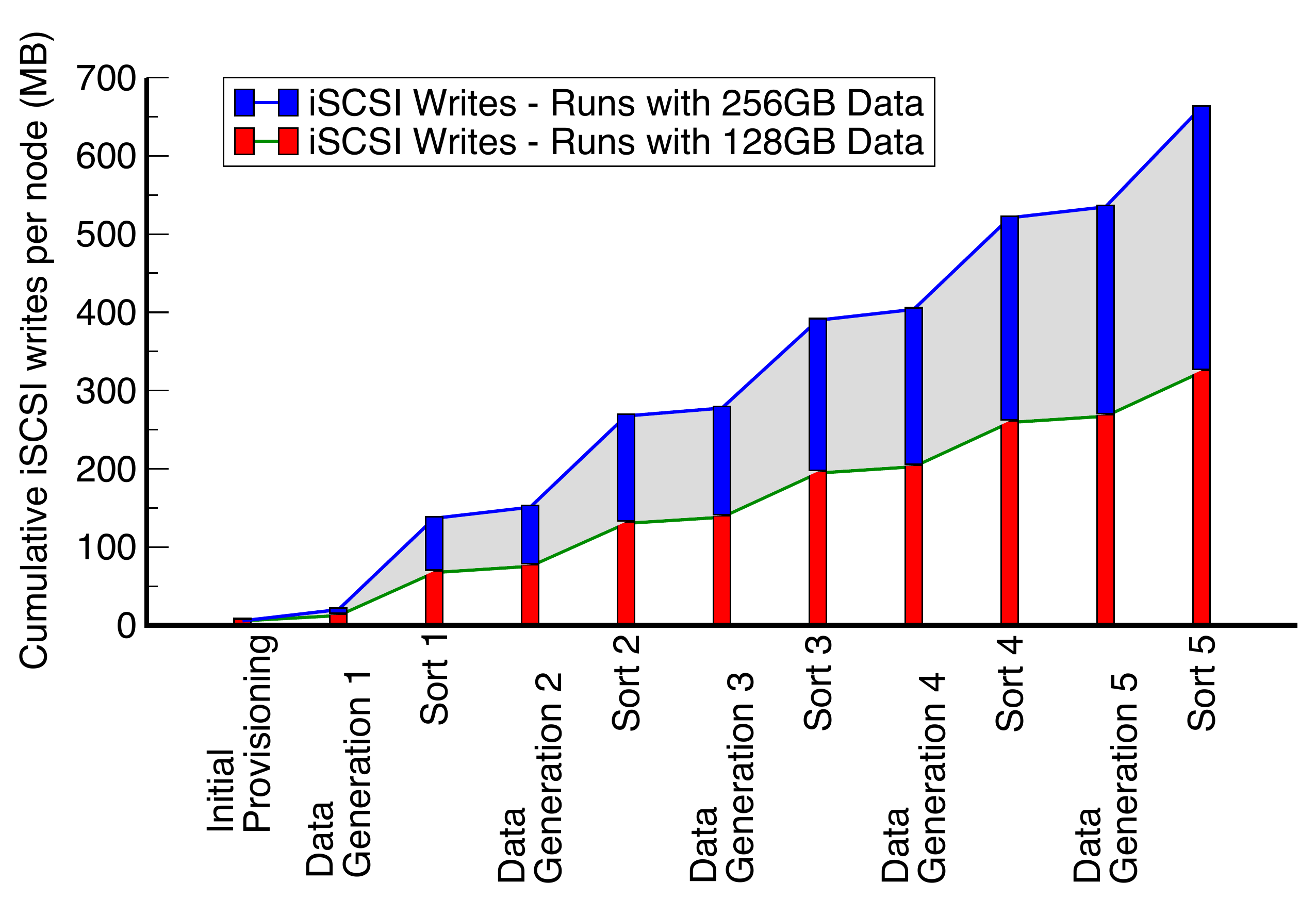}
		\caption{Write Traffic} \label{fig:write-traffic}
	\end{subfigure}
	\caption{Amount of read and write traffic passing through the \BMI\ iSCSI Service hosting the boot drive during provisioning of a Hadoop node and consecutive Hadoop application runs.}
  	\label{fig:nw-read-write-traffic}
\end{figure}

Figure~\ref{fig:nw-read-write-traffic} shows the per-node cumulative read and write traffic passing through the \BMI\ iSCSI Service during initial provisioning of a bare-metal Hadoop node and then over five consecutive ``data generation and sort'' jobs performed over the same node. ``Data generation and sort'' jobs of 128 GB and 256 GB are performed. Bare-metal nodes from the first environment are used during these experiments. The size of the image containing the operating system and the Hadoop packages was~8GB. Only the boot drive of the server is mounted remotely and the data drives are hosted on the local disk of the node in these experiments.

Figure~\ref{fig:read-traffic} shows that only $\sim$170MB of the $\sim$8GB image is read over the network during initial provisioning. Furthermore, both read and write curves flatten after repeated runs, demonstrating that (even with the 256GB case where the total data handled is substantially larger than the system memory) the file cache is effective at caching the boot drive. After initial boot and application start-up, the sustained read bandwidth incurred is around 3KB/s; effectively negligible.

Figure~\ref{fig:write-traffic} shows the writes to the network-mounted storage;  in contrast to the read case, log writes continue throughout the experiment, at an average rate of approximately 14~KB/s. On further examination, these writes target paths such as /var/log, /hadoop/log, and /var/run. (Note that in our deployments, /tmp and /swap are configured to reside on the local disk of servers.) Most of these writes are log file updates made by Hadoop. Although they could be directed to local storage, we did not do so due to their utility for debugging and negligible impact on the data rate.

\subsection{Performance of \BMI\ Provisioned Systems}
\label{sec:runtime-perf}

In this section we examine the impact of \BMI\ on the performance of applications and frameworks that generally run on bare-metal nodes. To this end, we compared the performance of these applications and frameworks when they run on top of \BMI-provisioned systems (network-mounted) and systems provisioned via Foreman (installed from a local disk).  

\begin{figure}[t]
\vspace{-1ex}
  \centering
    \includegraphics[width=0.48\textwidth]{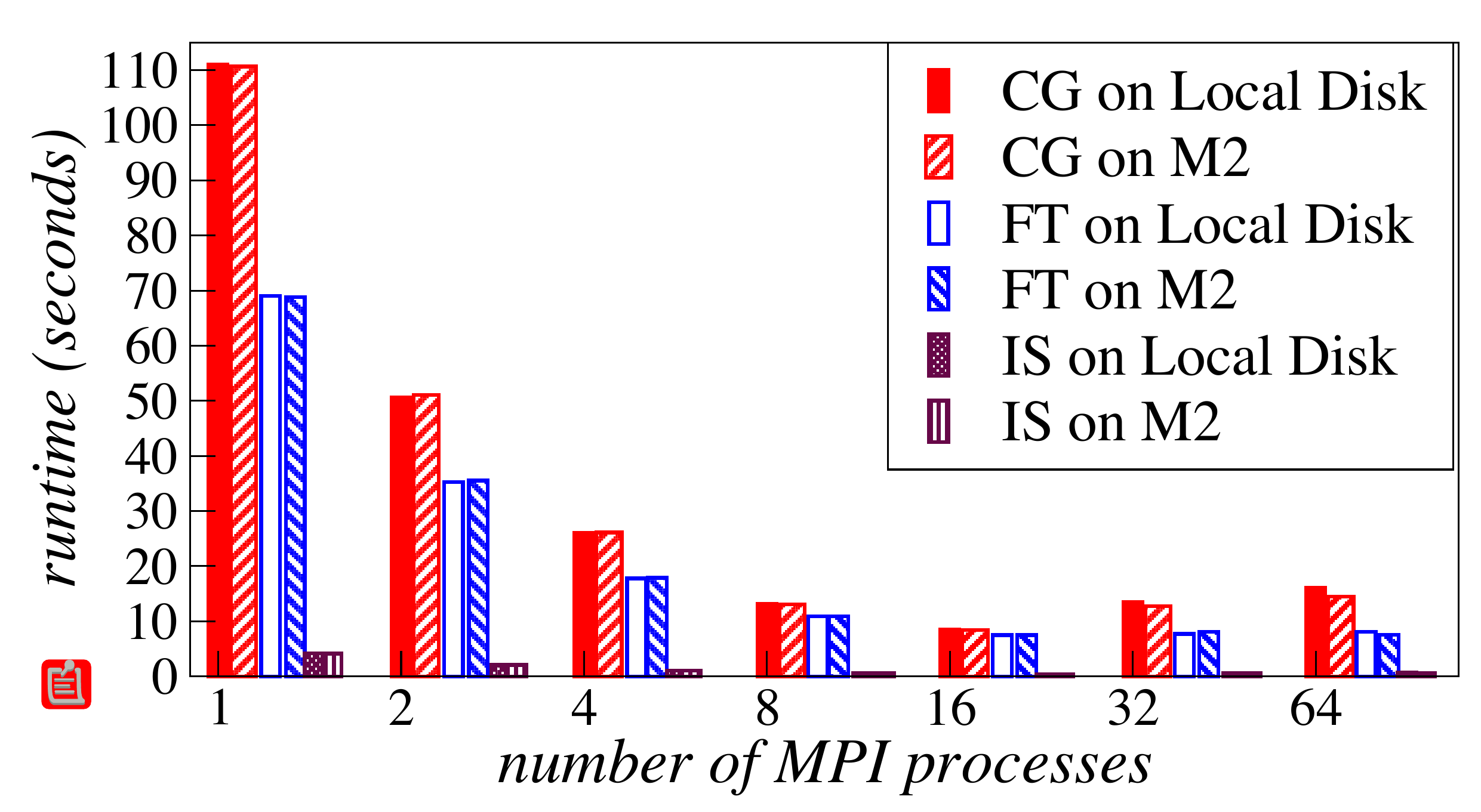}
    \caption{\BMI\ and local-disk runtime performance comparison of HPC applications (Conjugate-Gradient (CG), Fourier Transform (FT), and Integer Sort (IS) benchmarks from the NAS suite~\cite{bailey1992parallel}).} 
    \label{fig:HPC_perf}
\end{figure}

\subsubsection{HPC Applications Runtime Performance}

In Figure~\ref{fig:HPC_perf}, we compare the runtime of HPC applications (FT, CG, IS) from the NAS Parallel Benchmarks (NPB)~\cite{bailey1992parallel} running on Foreman-provisioned (installed from local disk) and \BMI\ provisioned (network-mounted) clusters.
We ran these benchmarks in our second environment.  
NPB is a set of programs designed to evaluate the performance of parallel supercomputers. Three benchmarks (i.e., FT, IS and CG) with distinct behaviors were used to evaluate the system. IS performs random memory access, CG has an irregular memory access and communication pattern, and FT does frequent all-to-all communications. We used class~B of the MPI version of NPB. Each benchmark was compiled to run with $2^n$ processes where $n \in \{1,\ldots, 8\}$. Each build was executed using OpenMPI on local and remote installations. 

As shown in Figure~\ref{fig:HPC_perf}, almost equal execution times were noted in the case of both \BMI\ and Foreman, resulting in same height bars. The results indicate that \BMI\ and diskless provisioning has no additional overheads when executing CPU- or memory-intensive HPC jobs. Note that it is already known that HPC applications perform well with remote boot drives as such solutions are frequently employed in Beowulf clusters and supercomputers. Hence good performance of \BMI\ is expected in this scenario.

\begin{figure}[t]
\vspace{-1ex}
  \centering
   \includegraphics[width=0.99\columnwidth]{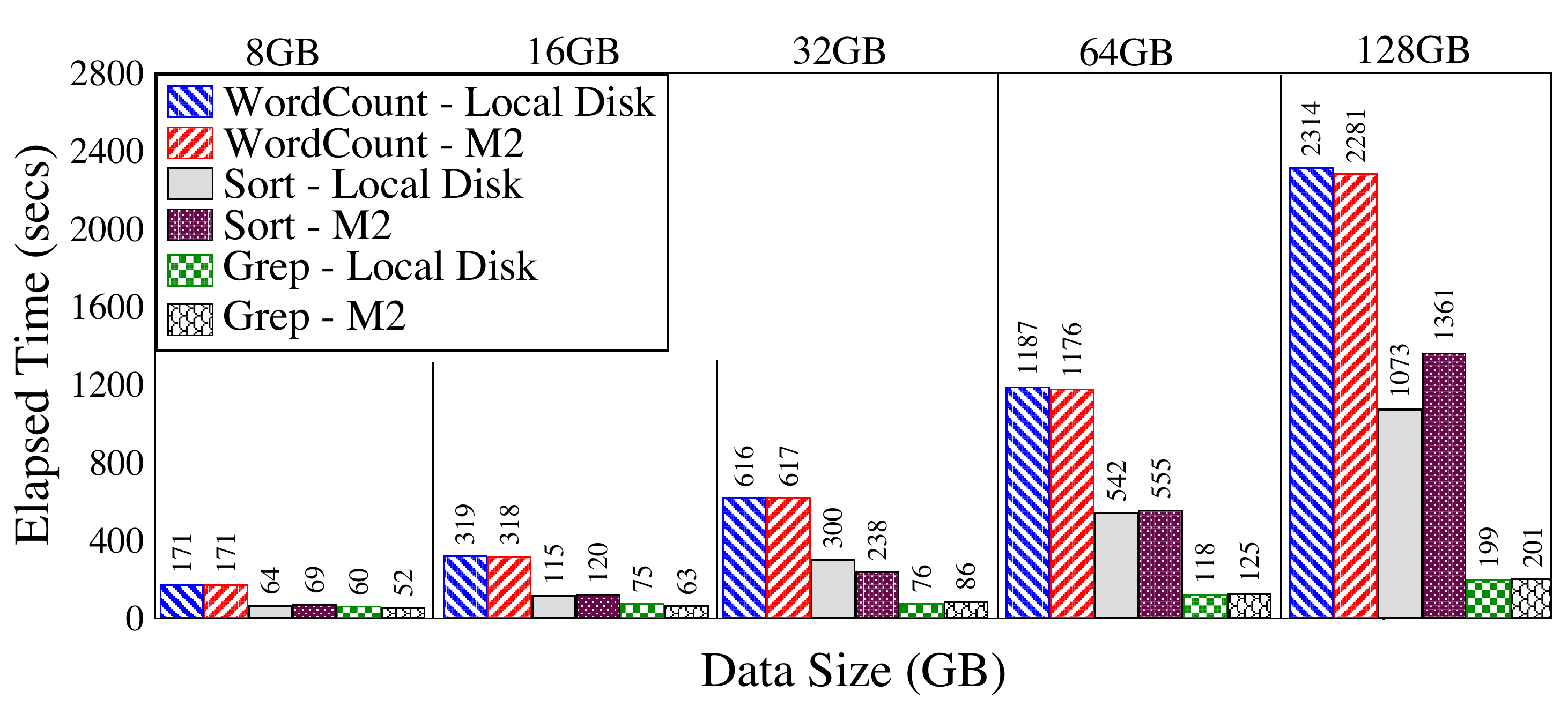}
  \vspace{-1ex}
  \caption{\BMI\ and local-disk runtime performance comparison of standard Hadoop benchmarks (WordCount, Sort, Grep).}
  \label{fig:hadoop-performance}
  \vspace{-1ex}
\end{figure}

\subsubsection{Hadoop Runtime Performance}
To measure \BMI's performance under high network and disk I/O usage we tested its performance when it runs Hadoop jobs. We performed a series of experiments on an 8-node Hadoop cluster as we varied the data set size between 8GB, 16GB, 32GB, 64GB and 128GB. We used the first environment for these experiments. 
Figure~\ref{fig:hadoop-performance} compares the runtime of standard Hadoop benchmarks (Sort, Grep, WordCount) running on clusters installed from local disks and from network-mounted clusters. In both cases, data disks hosting the Hadoop Distributed File System reside on local disks of the servers. Reported numbers are the average of five runs. We observe that deviations among runs on the same configuration are negligible. 

As shown in Figure~\ref{fig:hadoop-performance}, the difference in runtime performances of local-disk installed and \BMI\ provisioned systems are negligible, with the exception of the Sort experiments for 32GB data and 128GB data. We hypothesize that this exception may be caused by the non-deterministic behavior of random sorting benchmarks. The ``good'' performance of \BMI\ justifies our hypothesis that even for applications that create a significant amount of network traffic and disk I/O, the performance of the application does not get adversely impacted by remote mounting the boot drive.

\begin{figure}[t]
\vspace{-1ex}
  \centering
    \includegraphics[width=0.75\columnwidth]{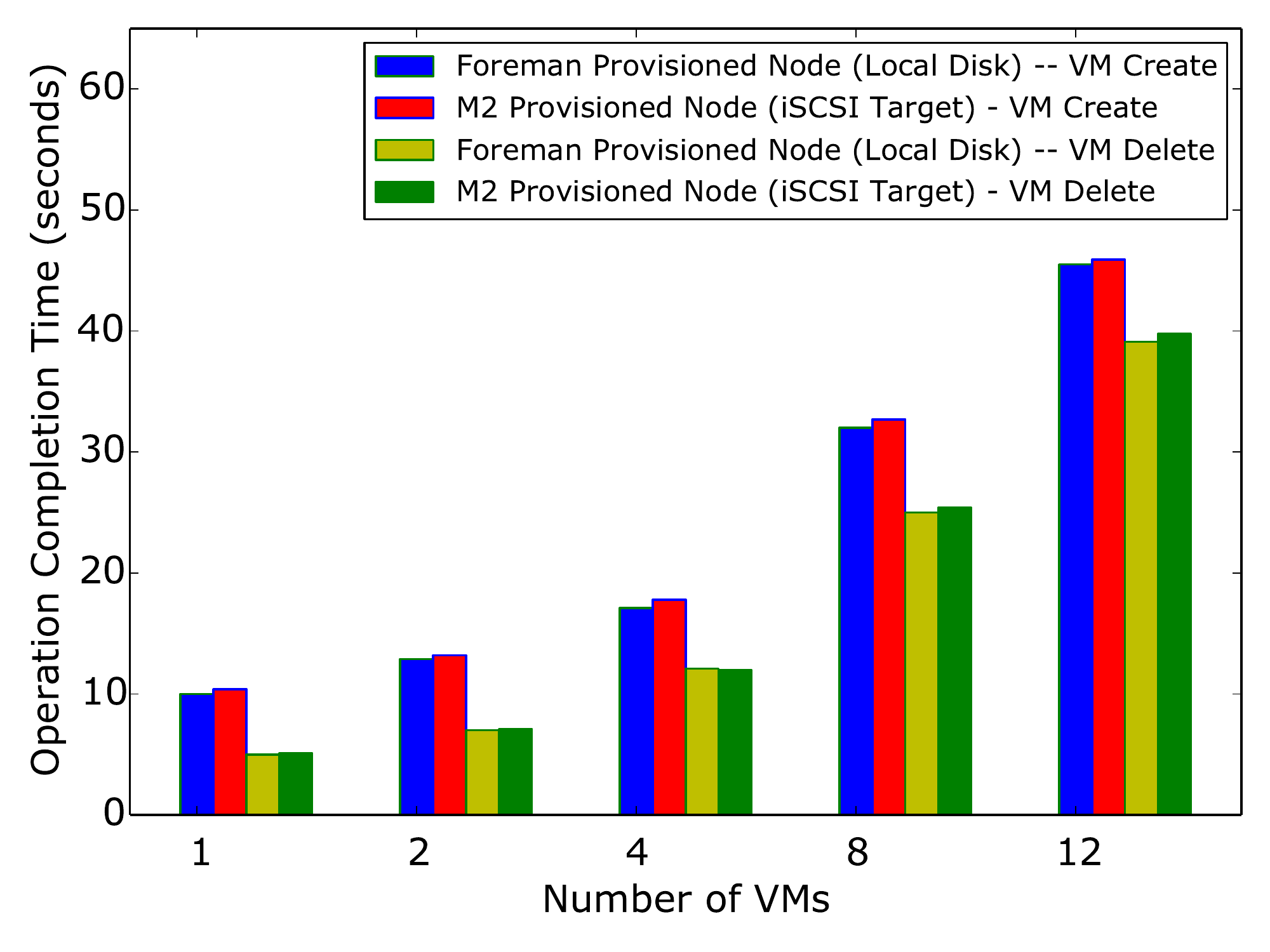}
      \vspace{-1ex}
        \caption{OpenStack operation performance comparison between Foremen and \BMI-provisioned nodes.}
    \label{fig:vm-create-delete}
      \vspace{-1ex}
\end{figure}

\subsubsection{OpenStack Operations Performance}
Cloud management systems such as OpenStack that offer virtualized services are also generally deployed on bare-metal servers. In this experiment, we set up OpenStack-based clouds to run on top of \BMI-provisioned and Foreman-provisioned systems in our second environment. We measured the performance of the two virtual machine operations, namely VM create and VM delete, in these two setups. We used the Rally benchmarking tool~\cite{rally} for these experiments, varying the number of parallel operation requests issued between 1 and~12. As shown in Figure~\ref{fig:vm-create-delete}, negligible performance degradation was observed for both creation and deletion operations between Foreman and \BMI-provisioned nodes.

\subsubsection{Latency and Throughput of Database Operations} 
\label{db_ops}
Due to their stringent performance requirements database systems are commonly deployed over bare-metal servers. To test if \BMI-provisioned servers can provide satisfactory performance while running database applications we compared the latency and throughput of various database operations when running commonly used databases on nodes provisioned via \BMI\ versus Foreman. Note that again, the data disks holding the actual database data is hosted on local disks in both cases.

\begin{figure}[t]
\vspace{-1ex}
  \centering
   \includegraphics[width=0.75\columnwidth]{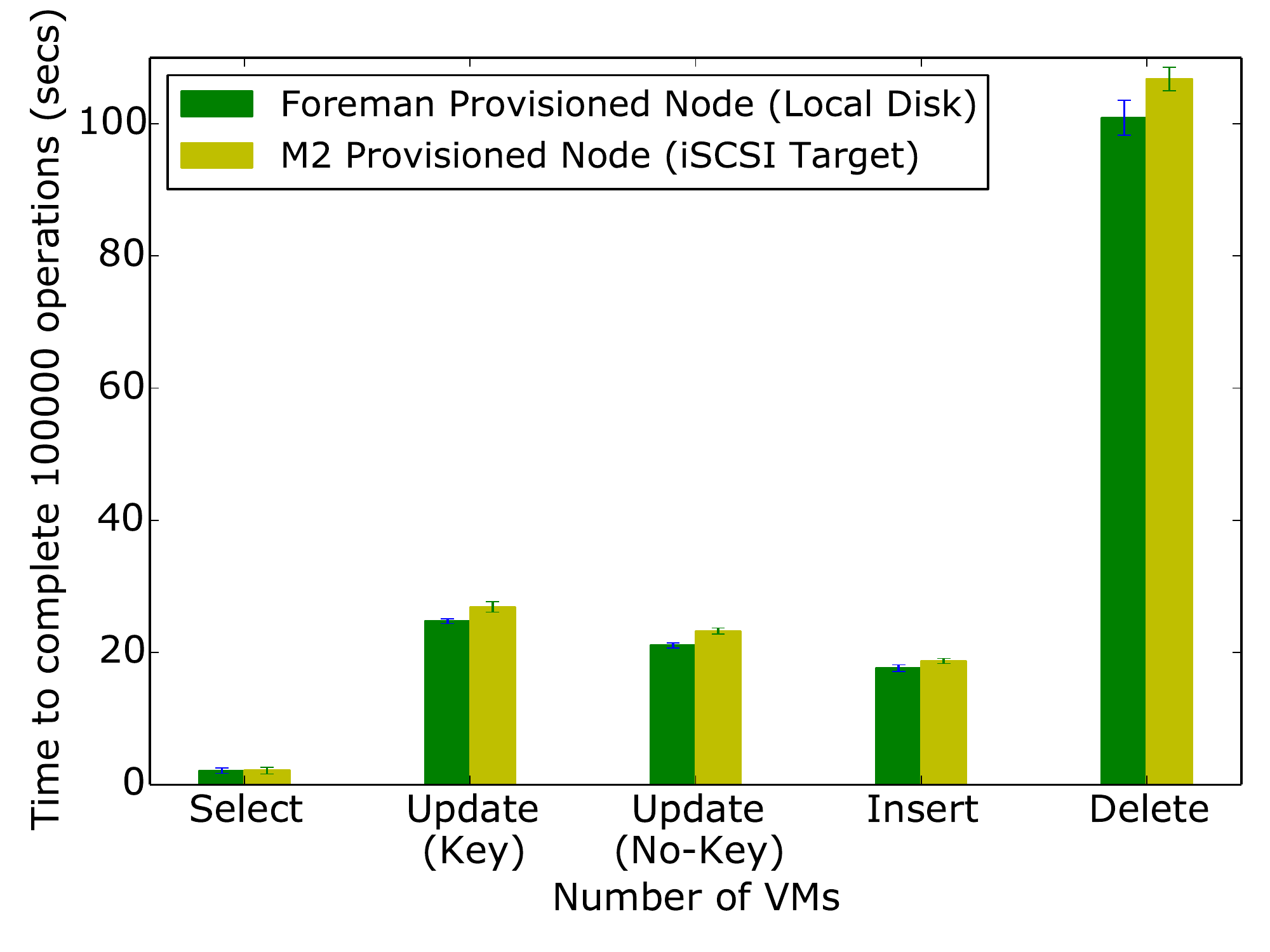}
  \vspace{-1ex}
  \caption{MariaDB operation latency comparison between Foremen and \BMI-provisioned nodes.}
  \label{fig:database-operations}
  \vspace{-1ex}
\end{figure}

\begin{figure}[t]
\vspace{-1ex}
  \centering
    \includegraphics[width=0.75\columnwidth]{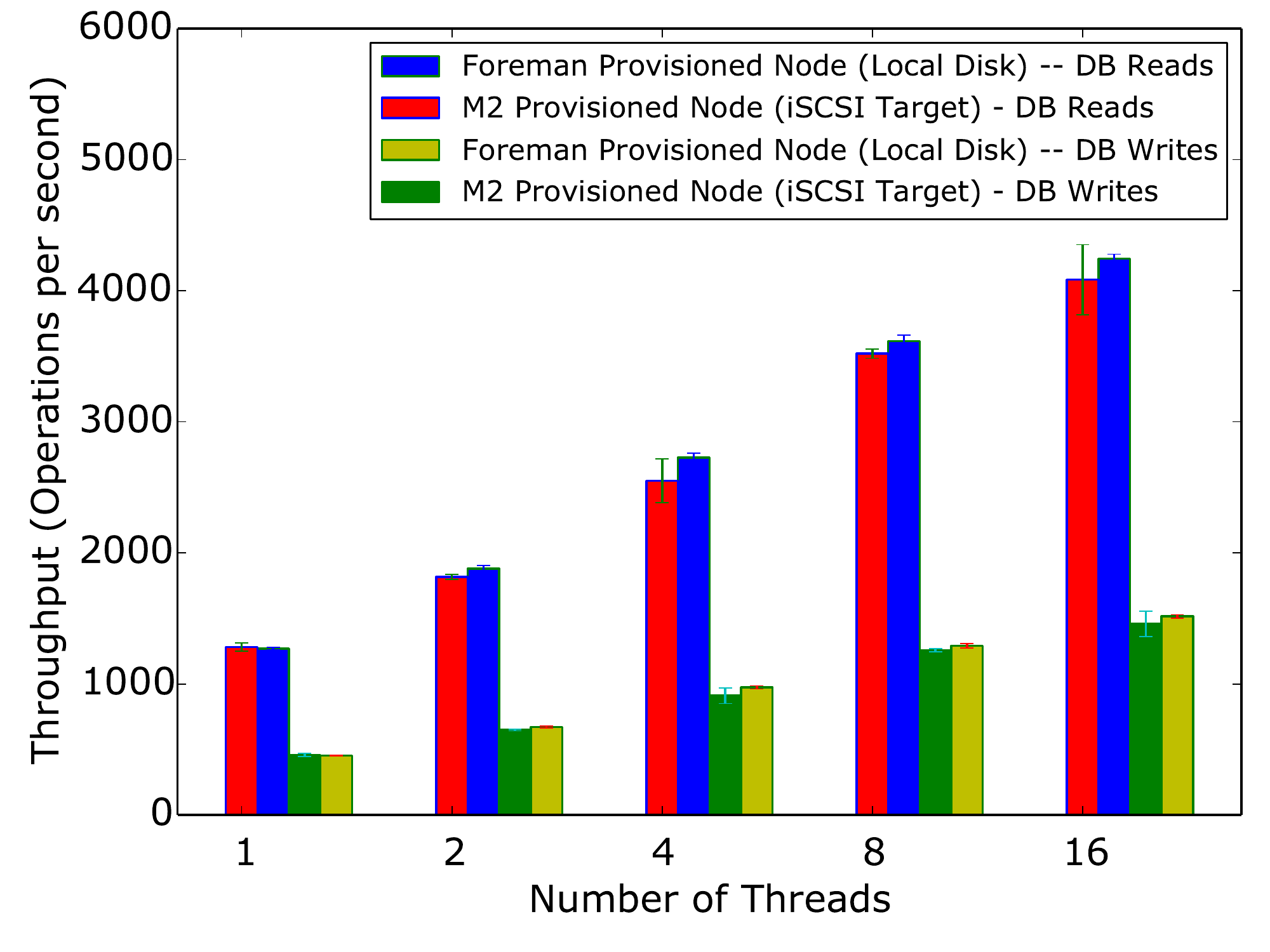}
      \vspace{-1ex}
        \caption{MySQL read/write throughput comparison between Foremen and \BMI-provisioned nodes.}
    \label{fig:db-read-write}
      \vspace{-1ex}
\end{figure}

Figure~\ref{fig:database-operations} compares the latency of database operations when running the popular MariaDB database on a node provisioned via \BMI\ versus Foreman. This experiment was performed using the Sysbench benchmarking tool~\cite{kopytov2004sysbench} in our second environment. Multi-threaded Online Transaction Processing (OLTP) tests for Select, Update, Insert and Delete operations were performed on the default ``sbtest''~\cite{kopytov2012sysbench} table generated by Sysbench with 1~million rows with InnoDB as the storage engine for MariaDB. The number of select operations executed during the test was 100,000, whereas 10,000 operations were executed for each of update, insert and delete operations. 
The update operation test had two versions --- updating an indexed column (Key) and updating a non-indexed column (No-key). For each test, the number of threads was fixed at~4. 

As seen in Figure~\ref{fig:database-operations}, there is a negligible impact on the latency of the select operation for MariaDB when running on a \BMI\ provisioned system. In contrast, in the case of update, insert and delete operations, we observe $\sim$4\% degradation in the case of \BMI\ provisioned nodes. 

Figure~\ref{fig:db-read-write} compares the MariaDB read and write throughput when it runs on a node provisioned via \BMI\ and Foreman. This experiment was also performed using Sysbench. In this experiment, we measured the total number of random reads and random writes performed in 300 seconds --- varying the number of threads. The throughput of both random reads and random writes in the case of \BMI-provisioned nodes was either at par with their Foreman-provisioned counterparts or saw a degradation less than~5\%. These experiments were also executed in our second environment.

Results in Figure~\ref{fig:database-operations} and Figure~\ref{fig:db-read-write} indicate that the impact of remote-mounting the boot drive to database performance is less than~5\%. This is potentially due to the excessive system memory use of the databases. Considering the additional benefits \BMI\ offers such as easy fault recovery and easy backup, we believe many deployments will find this additional impact tolerable.

\section{Future Work}
\label{sec:fw}

\subsection{Scaling}
\label{sec:future-scaling}

We believe that the evaluation shows the performance of \BMI\ to be sufficient at moderate scales and that the boot disk is not even necessarily a major factor for ongoing application performance. When \BMI\ does need to scale up, at least three strategies could be mixed and matched to do so.

\textbf{iSCSI multipath:} One of the advantages of selecting iSCSI as the gateway protocol is that many iSCSI clients support multipathing support. This means that clients can distribute queries across a number of iSCSI endpoints for both performance  and redundancy. When paired with a highy-scalable backend filesystem like Ceph or Lustre, this would mean that \BMI\ implementations could separately scale the iSCSI Service and the backend Storage Service.

\textbf{Caching:} Without breaking the consistency model, where the iSCSI Service is just a gateway and the backend filesystem is the coherence point, disks could be segmented into read-only and read-write components. For example, the /usr filesystem could be a read-only disk that is updated infrequently and /var could be on a read-write partition. Segmenting the disk this way would allow iSCSI Service to cache the read-only partitions locally, reducing load on the Storage Service overall and especially offloading the Storage Service in case of read-only hotspots.

\textbf{Custom hardware components:} Several vendors, like NetApp, Dell and EMC, sell high performance storage infrastructures that manage scaling and redundancy, and which expose that storage via an iSCSI endpoint. Due to the modular design of \BMI\ these commercial solutions cloud be employed as the storage service to improve scalability.

\subsection{Improved security}

Threats against \BMI\ can come from: the publicly-facing API service, the \BMI-provider itself or within an \BMI-serviced network; the cloud provider is trusted.
For threats against the API, \BMI\ relies on pluggable authentication modules.
In the case where tenants do not trust the cloud provider, they can maintain some confidentiality and integrity using encryption such as Linux LUKS~\cite{luks} or Windows Bitlocker~\cite{bitlocker}, though additional protections might be needed to prevent side channel analysis~\cite{SP:SteShi13} or block-level replay attacks.

In the current implementation, hostile nodes active on the network \BMI\ is managing can pose a threat due to the potential for accessing the boot disks of other nodes on the same network. This is in part due to the lack of strong identity inherent in several of the protocols \BMI\ uses. For example, one node could pretend to be another node booting, by spoofing its MAC address; a node could intercept the active iSCSI connection of another, since modern iSCSI implementations do not authenticate or encrypt every packet. Transport Layer Security (TLS)~\cite{rfc5246} or IPsec~\cite{rfc4301} could be applied to reduce this risk, but there still exists the bootstrapping problem: how does \BMI\ differentiate between a real node, and a compromised node that is faking its MAC or IP address?

To address this issue we are working on a solution that relies on Trusted Platform Modules~\cite{tpm}, which are tamper-resistant, discrete chips contained in some bare-metal nodes that give the node a cryptographic identity via a local public/private keypair.
A security-sensitive tenant could use the TPM as part of a protocol to grant cryptographically-guarded access to \BMI\ resources using attestation systems like \cite{keylime,Santos12policy-sealeddata}.
Such systems could also provide a defense against corrupted firmware.
We are working on this at present~\cite{secure-cloud}, though details of this work is out of scope for this paper.

\subsection{Transition between physical and virtual}

One interesting thought we had was: if \BMI 's goal is providing VM-like management capabilities for bare metal images, would it be possible to transition nodes between physical and \textit{virtual} nodes? This could be especially helpful for using the right amount of resources in HPC, dev/test or staging, where a lighter VM could be used for development and then a physical node for performance or predictability. Such a system could function by taking a Ceph RBD instance in use by a VM (a common choice in OpenStack clusters), then using \BMI\ to export that to a physical node via PXE and iSCSI. A preliminary test within our \BMI\ environment was able to do just this using a standard Linux distribution as the image. We are working to generalize this solution as a standard feature in the next \BMI\ release.

\balance
\section{Conclusion}
\label{sec:con}
In this work we proposed \BMI\ a system that brings the attractive image management capabilities (such as fast snapshotting, cloning, rapid provisioning etc.) of virtualized solutions to bare-metal systems. \BMI\ makes use of remote-mounted boot drives to host user images containing the operating system and applications and exploits advancements in disaggregated storage and networking technologies to offer high performance. Our analysis show that \BMI\ provisioned systems and frameworks perform as well as local-disk-based systems. We also show that rapid provisioning and snapshotting capabilities of \BMI\ unleash additional features and capabilities such as elasticity and support for fast transition among different frameworks for datacenter administrators.

\section{Acknowledgment}
We gratefully acknowledge Sourabh Bollapragada, Naved Ansari, Daniel Finn, Sirushti Murugesan and Paul Grosu for their significant contributions in development and documentation of \BMI. Also, Piyanai Saowarattitada, Chris Hill, Radoslav Nikiforov Milanov, Laura Kamfonik, Rahul Sharma, Rajul Kumar, and Sourabh Bollapragada for their assistance in the evaluations. 

Partial support for this work was provided by the MassTech Collaborative Research Matching Grant Program, National Science Foundation awards ACI-1440788, 1347525, 1149232 and 1414119 as well as the several commercial partners of the Massachusetts Open Cloud which include Brocade, Cisco, Intel, Lenovo, Red Hat, and Two Sigma.

\bibliographystyle{IEEEtran}

\bibliography{ms}

\end{document}